\documentclass[journal=jacsat,manuscript=article]{achemso}

\usepackage{xr}
\usepackage{chemformula} 
\usepackage[T1]{fontenc} 
\usepackage{makecell}
\usepackage{colortbl}
\usepackage[table]{xcolor}
\usepackage[labelfont=bf,labelsep=period]{caption}
\usepackage{siunitx}
\usepackage{chemformula}
\usepackage{lipsum}
\usepackage[upgreek, LGRgreek]{mathastext}

\author{Eugenio Gambari}
\affiliation[Sorbonne University]{Sorbonne Université, CNRS, Institut des Nanosciences de Paris, UMR7588, 4 place Jussieu, Paris, 75005, France}

\author{Hugo Le Du}
\affiliation[Sorbonne University]{Sorbonne Université, CNRS, Institut des Nanosciences de Paris, UMR7588, 4 place Jussieu, Paris, 75005, France}

\author{Mathieu Lizée}
\affiliation[Berlin University]{Fritz Haber Institute of the Max Planck Society,Berlin, Germany}
\alsoaffiliation[Berlin University]{Now at Fritz Haber Institute of the Max Planck Society,Berlin, Germany}

\author{Arindam Mukherjee}
\affiliation[Sorbonne University]
{Sorbonne Université, CNRS, Institut des Nanosciences de Paris, UMR7588, 4 place Jussieu, Paris, 75005, France}

\author{Laurent Limot}
\affiliation[Strasbourg University]{Université de Strasbourg, CNRS, Institut de Phsyique et Chime des Matériaux de Strasbourg, UMR7504, Strasbourg, F-67000, France}

\author{Fabrice Scheurer}
\affiliation[Strasbourg University]{Université de Strasbourg, CNRS, Institut de Phsyique et Chime des Matériaux de Strasbourg, UMR7504, Strasbourg, F-67000, France}

\author{Marie D'angelo}
\affiliation[Sorbonne University]{Sorbonne Université, CNRS, Institut des Nanosciences de Paris, UMR7588, 4 place Jussieu, Paris, 75005, France}

\author{François Debontridder}
\affiliation[Sorbonne University]{Sorbonne Université, CNRS, Institut des Nanosciences de Paris, UMR7588, 4 place Jussieu, Paris, 75005, France}

\author{Tristan Cren}
\affiliation[Sorbonne University]{Sorbonne Université, CNRS, Institut des Nanosciences de Paris, UMR7588, 4 place Jussieu, Paris, 75005, France}

\author{Marie Hervé}
\email{marie.herve@sorbonne-universite.fr}
\affiliation[Sorbonne University]{Sorbonne Université, CNRS, Institut des Nanosciences de Paris, UMR7588, 4 place Jussieu, Paris, 75005, France}

\title[Moiré pattern multiplicity driven by electronic effects in two-dimensional \ch{CrCl3}/Au heterostructures ]{Moiré pattern multiplicity driven by electronic effects in two-dimensional \ch{CrCl3}/Au heterostructures.}

\abbreviations{IR,NMR,UV}

\begin{document}


\begin{abstract}
    Moiré patterns are a central motif in van der Waals heterostructures arising from the superposition of two-dimensional (2D) incommensurate lattices. These patterns reveal a wealth of correlated effects, influencing electronic, magnetic, and structural phenomena. While diffraction techniques typically resolve multiple moiré wave-vectors corresponding to the incommensurate nature of the underlying lattices, Scanning Tunneling Microscopy (STM) often reveals only a dominant superperiod. In this work, we address this apparent discrepancy through an STM study of a twisted monolayer of \ch{CrCl3} on Au(111). We observe the coexistence of several moiré patterns at a fixed twist angle, whose relative intensity depends on the tunneling bias. Fourier analysis of STM data uncovers hidden higher-order moiré components not visible in STM topographic images, while spectroscopy maps reveal that the spectral weight of each pattern varies with electron energy. Our results establish that STM selectively probes on the same area distinct moiré modulations depending on electronic confinement, providing a unified framework that reconciles real space and reciprocal space observations of complex moiré superstructures. 
\end{abstract}

\section{Introduction}

Moiré patterns in  two-dimensional materials are ubiquitous. They are found in van der Waals heterostructures such as Gr/Ir, Gr/Pt \cite{zeller_what_2014,loginova_defects_2009,blanc_local_2012, ndiaye_structure_2008,merino_strain-driven_2011}, \ch{MoS2}/Au \cite{krane_moire_2018,reidy_direct_2021} when the lattice parameters of the two materials are incommensurate, or in twisted bilayers of van der Waals materials such as graphene, \ch{CrI3}, or \ch{WSe2} \cite{gambari_higher_2024,merino_strain-driven_2011,li_observation_2010,cao_unconventional_2018,xie_evidence_2023,cheng_electrically_2023,song_direct_2021,guo_superconductivity_2025,foutty_mapping_2024,li_imaging_2021}. For specific orientations, moiré patterns can lead to the emergence of flat bands \cite{balents_superconductivity_2020,li_imaging_2021-1} giving rise to many exciting physical phenomena. Recently, superconductivity has been discovered in materials that are not intrinsically superconducting, such as twisted bilayer graphene or \ch{WSe2} \cite{cao_unconventional_2018,guo_superconductivity_2025}. Wigner crystals have been observed in moiré lattices of \ch{WSe2}/\ch{WS2} \cite{li_imaging_2021}. Non-collinear magnetic orders have been reported in materials in which they are usually collinear, such as in twisted bilayers of \ch{CrI3} \cite{xie_evidence_2023,cheng_electrically_2023,song_direct_2021}. 

Moiré patterns are a direct manifestation of the quasiperiodic structure induced by the superposition of two incommensurate periodic lattices. In reciprocal space this leads to an infinite number of satellite peaks surrounding the Bragg peaks, as observed by diffraction techniques \cite{loginova_defects_2009, langer_sheet_2011, reidy_direct_2021}. However, other experimental probes accessing to the real space order, such as scanning probe microscopy techniques, tend to reveal a single dominant moiré pattern that is commonly interpreted as a commensurate superperiod \cite{artaud_universal_2016,yang_moire_2022,pham_higher-indexed_2022,gunther_method_2021,hennighausen_twistronics_2021,schwab_skyrmion_2025}. 

The use of a periodic superlattice description in \textit{ab-initio} calculations is also quite common. For example this is used for the prediction of a periodic non-collinear magnetic order in  twisted bilayers of \ch{CrI3} \cite{fumega_moire-driven_2023}. It sometimes happens that the same system studied by diffraction and STM leads to contradictory observations, as in the case of \ch{MoS2}/Au \cite{reidy_direct_2021}. The two approaches, in real and reciprocal space, seem to differ conceptually. In this paper, we link them through a real space study exhibiting the coexistence of several moiré patterns in a same location. In particular, our analysis reveals that STM is selectively sensitive to different coexisting moiré patterns, depending on the energy of the probed electrons. 

When examining a moiré pattern arising from the superposition of two 2-dimensional periodic lattices, as observed in van der Waals heterostructures, the mathematical representation of the moiré pattern in real space can become quite complex, especially when higher-order moiré patterns are involved \cite{zeller_what_2014, gambari_higher_2024, zeller_scalable_2012, blanc_local_2012, langer_sheet_2011, loginova_defects_2009, merino_strain-driven_2011}. In contrast, the description in reciprocal space is quite straightforward \cite{zeller_what_2014, ster_moire_2019, gambari_higher_2024}, a moiré pattern is generated when two periodic functions with spatial frequencies $\boldsymbol{Q}_1$ and $\boldsymbol{Q}_2$ are multiplied together. This yields a beating frequency $\boldsymbol{Q}_\text{moiré}=\boldsymbol{Q}_{1}-\boldsymbol{Q}_{2}$, corresponding to the moiré wave vector. For two 2-dimensional atomic lattices, the general reciprocal lattice vectors of lattice 1 and 2, $\boldsymbol{Q}_{1}^{i,j}$ and $\boldsymbol{Q}_{2}^{k,l}$ can be expressed as linear combinations of the primitive reciprocal lattice vectors $\boldsymbol{Q}_{1}^{1,0}$, $\boldsymbol{Q}_{1}^{0,1}$,  $\boldsymbol{Q}_{2}^{1,0}$ and $\boldsymbol{Q}_{2}^{0,1}$, as:
\vspace{-10mm}

\begin{align}
\boldsymbol{Q}_{1}^{i,j} &=i\boldsymbol{Q}_{1}^{1,0}+j\boldsymbol{Q}_{1}^{0,1}   &   \boldsymbol{Q}_{2}^{k,l} &=k\boldsymbol{Q}_{2}^{1,0}+l\boldsymbol{Q}_{2}^{0,1}   
\end{align}

This leads to an infinite number of possible Fourier components of the moiré patterns, where the wave vectors can be expressed as:

\begin{equation}
    \boldsymbol{Q}_\text{moiré}^{i,j,k,l}=\boldsymbol{Q}_{1}^{i,j}-\boldsymbol{Q}_{2}^{k,l}
\end{equation}

Electron diffraction experiments often reveal multiple moiré patterns in the Fourier space for a given specific orientation between the atomic lattices \cite{loginova_defects_2009, langer_sheet_2011, reidy_direct_2021} whereas STM experiments often identify a single dominant moiré pattern \cite{reidy_direct_2021, loginova_defects_2009}. A possible reason for that could be that electron diffraction integrate over large surface areas while STM reveal a local order. Pham et al. \cite{pham_higher-indexed_2022} showed that the moiré contrast observed in STM topographs depends on the tunneling bias, suggesting that the modulation is electronic rather than structural. However, STM topography alone are not straighforward to interpret in real space since several distinct moiré modulations may coexist at the same energy and within the same real-space region, with often one dominant component that hides less intense modulations. We will show that an efficient way to reveal the coexistence of multiple moiré patterns consist in combining energy-resolved spectroscopy with Fourier-space analysis. Here, we study a \ch{CrCl3} monolayer on Au(111) and show that the moiré landscape is far richer than what topography alone reveals. Fourier analysis of STM data uncovers hidden higher-order moiré components not visible in STM topographic images, while spectroscopy maps reveal that the spectral weight of each pattern varies with electron energy. Our results establish that STM selectively probes on a same area distinct moiré modulations depending on electronic confinement. These findings provide a unified framework for complex moiré superstructures, and, more generally, imply that the emergent electronic, magnetic, and correlated properties of moiré materials cannot be fully understood within a single-period description.

\section{Results and discussion}

\begin{figure}[ht]
    \centering
    \includegraphics[width=1\linewidth]{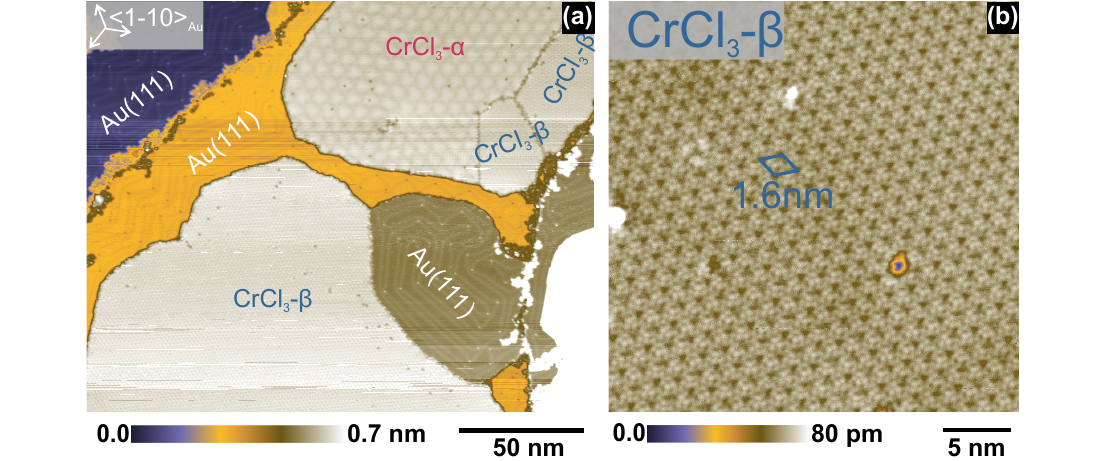}
    \caption{STM topography of \ch{CrCl3} islands deposited on Au(111). \textbf{(a)} Large scale STM topography displaying the two types of \ch{CrCl3} islands: $\ch{CrCl3}^{\alpha}$ and $\ch{CrCl3}^{\beta}$. Their atomic lattices are twisted by \ang{30} with respect to each other and their surfaces show two distinct moiré patterns. \textbf{(b)} STM topography of a $\ch{CrCl3}^{\beta}$ island, the moiré pattern has a periodicity of \qty{1.6}{\nm}. Tunneling parameters: (a) $V = \qty{1.2}{\volt}$, $I = \qty{100}{\pico\ampere}$; (b) $V = \qty{1.4}{\volt}$, $I = \qty{100}{\pico\ampere}$.}\label{fig1}
\end{figure}

\begin{figure}[H]
    \centering
    \includegraphics[width=0.8\linewidth]{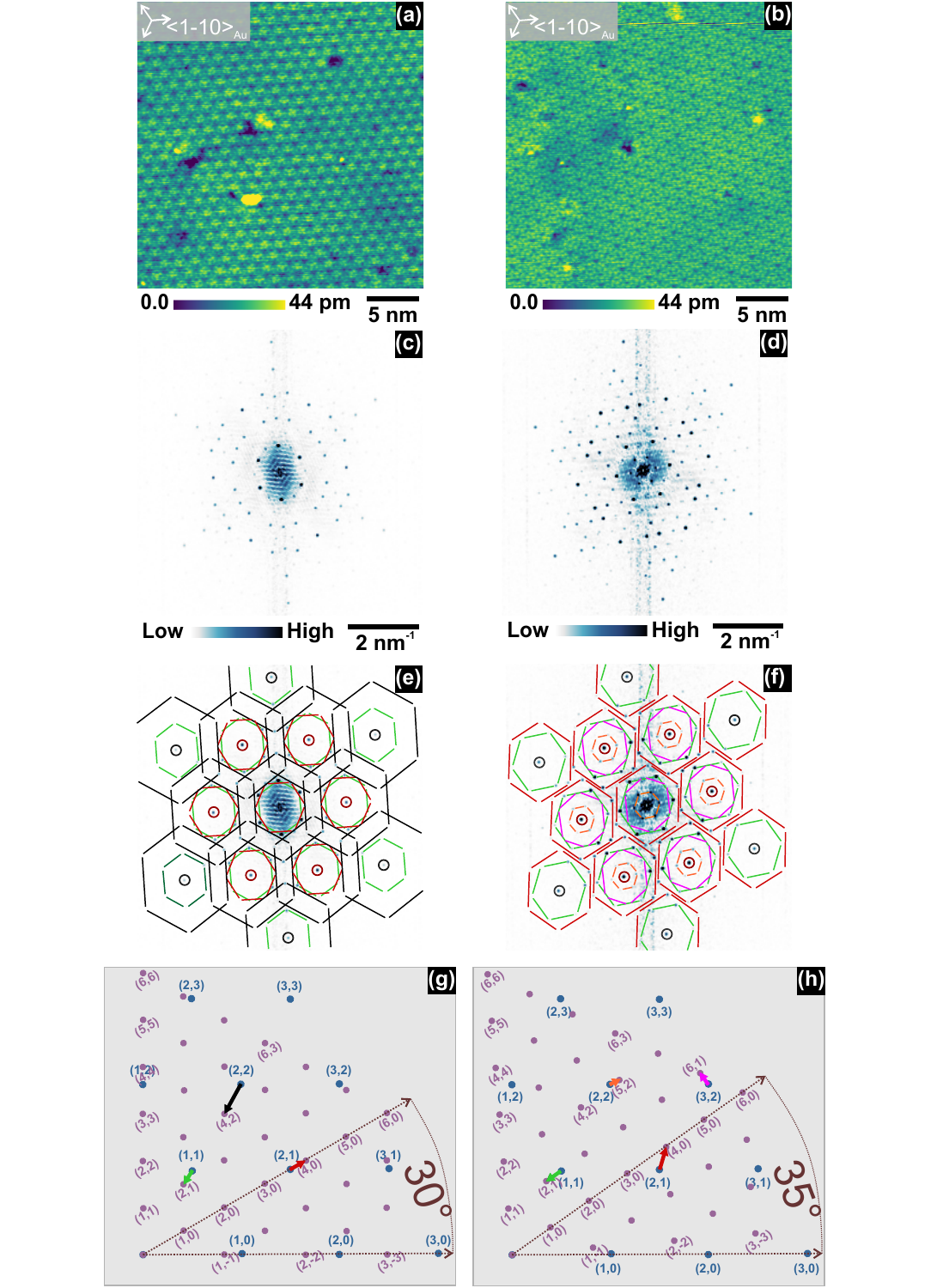}
    \caption{Analysis of multiple moiré patterns observed in \ch{CrCl3}$^{\beta}$ islands. \textbf{(a)},\textbf{(b)} STM topographies of $\beta$ islands with atomic lattices rotated by \ang{30} and \ang{35} relative to the <1$\overline{1}$0>$_\text{Au}$ direction. Tunneling parameters: $V = \qty{1.5}{\volt}$, $I = \qty{100}{\pico\ampere}$. \textbf{(c)},\textbf{(d)} Corresponding Fourier transform images. \textbf{(e)},\textbf{(f)} Same FT images as in (c),(d), with highlighted Bragg spots: the \ch{CrCl3} spots are marked by red and black circles, while the moiré spots are located at the vertices of colored hexagons. \textbf{(g)},\textbf{(h)} Graphical constructions of the \ch{CrCl3} (violet) and Au (blue) reciprocal lattices for twist angles of \ang{30} and \ang{35}. Colored arrows indicate the Bragg spot pairs responsible for the distinct moiré patterns.}
    \label{fig2}
\end{figure}

\ch{CrCl3} monolayer islands were grown on a clean Au(111) substrate by molecular beam epitaxy (MBE) \cite{gambari_higher_2024}. Figure~\ref{fig1}a displays a large-scale STM topography of the surface. The bare atomic terraces of the Au(111) substrate are partially covered by \ch{CrCl3} monolayer islands, presenting a width larger than \qty{100}{\nm}. These islands can grow with various crystalline orientations with respect to the Au(111) surface, leading to the formation of different moiré patterns (see Supporting Information S1). We can sort the island orientations into two main sets, labeled as $\alpha$ and $\beta$. The dense crystallographic directions of the \ch{CrCl3}$^{\alpha}$ islands (Figure~\ref{fig1}a) are aligned with those of the underlying Au(111) lattice (<1$\overline{1}$0>$_\text{Au}$), while the \ch{CrCl3}$^{\beta}$ islands (Figure~\ref{fig1}b) are twisted by about \ang{30}. In the \ch{CrCl3}$^{\alpha}$ islands, a dominant moiré pattern with a periodicity of \qty{6.2}{\nm} is observed (Figure~\ref{fig1}a). By contrast, a much shorter periodicity of approximately \qty{1.6}{\nm} is observed in \ch{CrCl3}$^{\beta}$ islands (Figure~\ref{fig1}b). In a prior study, we conducted a detailed analysis of the moiré pattern and the associated topological defects found in the $\ch{CrCl3}^{\alpha}$ islands \cite{gambari_higher_2024}. In the present work, we focus on the $\ch{CrCl3}^{\beta}$ islands, showing that the observed superstructure stems from the superposition of multiple moiré patterns. Figure~\ref{fig2} shows a detailed analysis of the moiré patterns in $\ch{CrCl3}^{\beta}$ islands. The crystallographic orientation of these domains varies slightly from island to island, between about \ang{30} and \ang{35} relative to the <1$\overline{1}$0>$_\text{Au}$ direction. Two limiting cases are shown in Figures~\ref{fig2}a and~\ref{fig2}b: one island showing a \ang{30} twist and the other a \ang{35} one. For these configurations, as well as for intermediate angles, the observed superstructures arise from the superposition of multiple moiré patterns, as revealed by reciprocal space analysis. Figures~\ref{fig2}c and \ref{fig2}d are obtained by performing the 2D Fourier Transform (FT) of the STM topographies in Figures~\ref{fig2}a and \ref{fig2}b, respectively. 

These FT images reveal numerous spots. Figures~\ref{fig2}e and~\ref{fig2}f present the same Fourier-transform images as in Figures~\ref{fig2}c and~\ref{fig2}d, with Bragg spots and moiré satellites highlighted. Red circles indicate the main Bragg spots corresponding to the dense atomic directions of \ch{CrCl3}, while black circles denote higher-order Bragg spots rotated by \ang{30} relative to the dense atomic axes. The remaining spots correspond to satellites from multiple moiré patterns. The superposition of the \ch{CrCl3} and Au(111) hexagonal lattices generates a hexagonal moiré superlattice in real space, which manifests in reciprocal space as six satellite spots forming a hexagon around each \ch{CrCl3} Bragg peak. In Figure~\ref{fig2}e, where the \ch{CrCl3} lattice is twisted by \ang{30} relative to <1$\overline{1}$0>$_\text{Au}$, three moiré patterns (highlighted by green, red, and black hexagons) are visible both at the Brillouin zone center and around the Bragg peaks. For a domain twisted by \ang{35} (Figure~\ref{fig2}f), four distinct moiré patterns appear, marked by orange, pink, green, and red hexagons. Each vertex of the moiré hexagons in the FT images originates from the conjunction of one \ch{CrCl3} and one Au(111) Bragg spot, denoted ${(i,j)}_{\ch{CrCl3}}$ and ${(k,l)}_{\ch{Au}}$. To identify which Bragg spots generate a given moiré component, we superimposed the reciprocal lattices of \ch{CrCl3} (violet) and Au(111) (blue), as shown in Figures~\ref{fig2}g and~\ref{fig2}h. In Figure~\ref{fig2}g, the \ch{CrCl3} lattice is twisted by \ang{30} relative to <1$\overline{1}$0>$_\text{Au}$, and by \ang{35} in Figure~\ref{fig2}h. This angular relationship is illustrated by the brown dashed arrows indicating the $\boldsymbol{Q}_{\ch{Au}}^{1,0}$ and $\boldsymbol{Q}_{\ch{CrCl3}}^{1,0}$ directions. For clarity, only one quadrant of reciprocal space is shown; the complete construction is provided in the Supporting Information S2. By measuring the size and orientation of a given $\boldsymbol{Q}_\text{moiré}$ in Figures~\ref{fig2}e and~\ref{fig2}f, we can determine which pair of Bragg vectors, from Au and \ch{CrCl3}, is involved. As highlighted by green hexagons  in Figures~\ref{fig2}e and~\ref{fig2}f, the dominant moiré pattern arises from the same combination of Au(111) and \ch{CrCl3} lattice vectors, although its size and orientation differ between the two domains. In the \ang{30} twisted domain (Figure~\ref{fig2}e), the green highlighted moiré spots are aligned with the higher order \ch{CrCl3} Bragg spots (marked by black circles). The associated $\boldsymbol{Q}_\text{moiré}$ is shown as a green arrow in Figure~\ref{fig2}g and is oriented at \ang{30} from $\boldsymbol{Q}_{\ch{CrCl3}}^{1,0}$. In the \ang{35} domain (Figure~\ref{fig2}f), the same moiré spots line up with the first order Bragg spots marked by the red circles, and the corresponding $\boldsymbol{Q}_\text{moiré}$ (the green arrow in Figure~\ref{fig2}h) lies parallel to $\boldsymbol{Q}_{\ch{CrCl3}}^{1,0}$. We found that the dominant moiré present in these twisted domains correspond to the conjunction of the (2,1)$_{\ch{CrCl3}}$ and (1,1)$_{\ch{Au}}$ spots. The graphical constructions in Figures~\ref{fig2}e and~\ref{fig2}f illustrate that a $5^\circ$ twist of the \ch{CrCl3} lattice results in an approximately \ang{30} rotation of the two moiré patterns. By analyzing in this manner all moiré patterns observed in the \ang{30} and \ang{35} twisted domains, we identified three and four pairs of Bragg spots, respectively, as indicated by the colored arrows in Figures~\ref{fig2}g and~\ref{fig2}h. The corresponding Bragg spots are listed in Table~\ref{table1}.

\newpage

{\renewcommand{\arraystretch}{1.4}
\begin{table}[H]
    \centering
    \footnotesize
    \resizebox{\textwidth}{!}{
    \begin{tabular}{|c c c c c c|}
    \hline
    \rowcolor{cyan!10}
        \textbf{Twist angle} & \makecell[c]{\textbf{Green}\\(2,1)$_{CrCl_3}$/(1,1)$_{Au}$} & \makecell[c]{\textbf{Black}\\(2,1)$_{CrCl_3}$/(1,1)$_{Au}$} & \makecell[c]{\textbf{Red}\\(4,0)$_{CrCl_3}$/(2,1)$_{Au}$}& \makecell[c]{\textbf{Orange}\\(5,2)$_{CrCl_3}$/(2,2)$_{Au}$}& \makecell[c]{\textbf{Pink}\\(6,1)$_{CrCl_3}$/(3,2)$_{Au}$}\\[2ex]
    
        \ang{30}& Detected& Detected& Detected& Not detected& Not detected\\[1ex]
        \hline
        \ang{35}& Detected& Not detected& Detected& Detected& Detected\\[1ex]
        \hline
    \end{tabular}}
    \caption{Summary of \ch{CrCl3} and Au Bragg spot pairs generating different moiré components in \ang{30} and \ang{35} twisted domains.}\label{table1}
\end{table}}

Slight variations in the orientation of \ch{CrCl3} lead to the appearance or disappearance of specific moiré patterns. For example, the moiré marked in black in Figure~\ref{fig2}e, resulting from the conjunction of (4,2)$_{\ch{CrCl3}}$ and (2,2)$_\text{Au}$, disappears when the \ch{CrCl3} lattice is twisted from \ang{30} to \ang{35}. In the \ang{35} domain (Figure~\ref{fig2}f), two additional moiré patterns emerge (in orange and pink)  and originate from the pairs (5,2)$_{CrCl_3}$/(2,2)$_{Au}$ and (6,1)$_{CrCl_3}$/(3,2)$_{Au}$ respectively. For both orientations of \ch{CrCl3} presented in Figure~\ref{fig2}, the various overlapping moiré patterns have similar periodicities, making them challenging to distinguish in real space images. The topographic contrast in Figures~\ref{fig2}a and~\ref{fig2}b is dominated by a moiré stemming from the (2,1)$_{CrCl_3}$ and (1,1)$_{Au}$ conjunction (the green arrows in Figures~\ref{fig2}g and~\ref{fig2}h). For these two twist angles, disentangling the contributions of overlapping moiré patterns requires reciprocal space analysis, as shown in Figures~\ref{fig2}e and~\ref{fig2}f.

\begin{figure}[ht]
    \centering
    \includegraphics[width=1\linewidth]{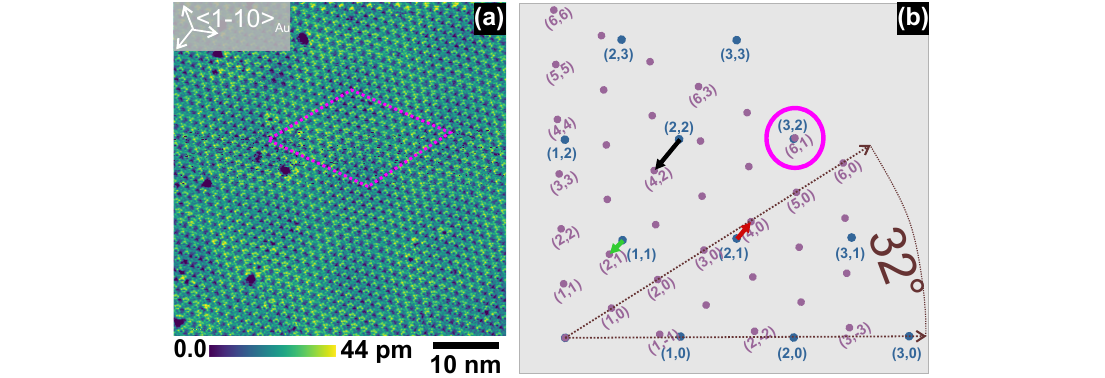}
    \caption{\textbf{(a)} STM topography of a $\ch{CrCl3}^{\beta}$ island twisted of approximately \ang{32} relative to the <1$\overline{1}$0>$_\text{Au}$ direction, showing a weak, long-wavelength moiré modulation ($\sim$\qty{20}{\nm}) superimposed on the dominant $\sim$\qty{1.6}{\nm} moiré pattern. Tunneling parameters: $V = \qty{1.2}{\volt}$, $I = \qty{100}{\pico\ampere}$.
    \textbf{(b)} Graphical construction of \ch{CrCl3} and Au reciprocal lattices. The Bragg spot pair generating the longer period moiré is circled in pink.}\label{fig3}
\end{figure}

At intermediate twist angles, such as $32^\circ$, the Bragg spots associated with the pink moiré, (6,1)$_{\ch{CrCl3}}$ and (3,2)$_{\ch{Au}}$, approach one another in reciprocal space (see Figure~\ref{fig3}b). As a consequence, the corresponding moiré wave vector becomes so small that its spots are buried within the central region of the FT, where background contributions from other electronic structure effects dominate. In real space, this moiré manifests as a long-wavelength modulation, clearly distinguishable from the shorter-period moirés, despite its comparatively weak spectral weight. Figure~\ref{fig3}a shows an STM topography of a region where the \ch{CrCl3} lattice is twisted by approximately \ang{32}. Alongside the main moiré modulation, with a real-space period of about \qty{1.6}{\nm}, a fainter superstructure with a larger period of approximately \qty{20}{\nm} is also observed.

\begin{figure}[ht]
    \centering
    \includegraphics[width=1\linewidth]{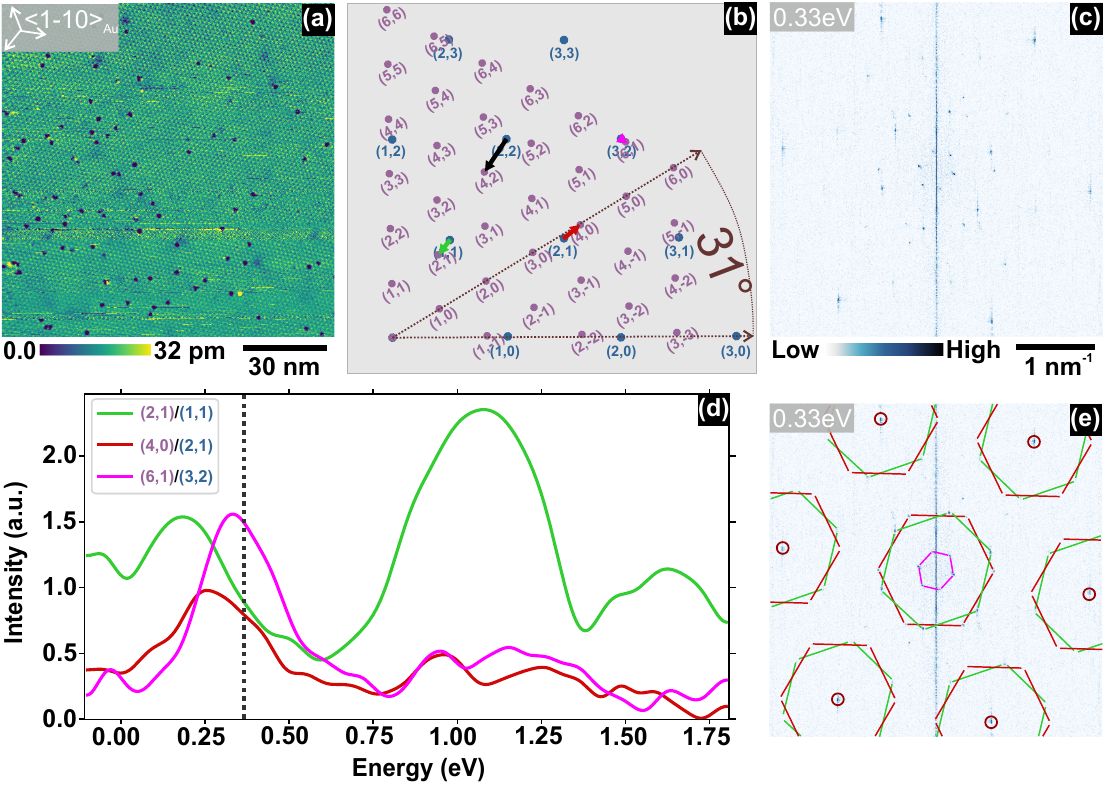}
    \caption{\textbf{(a)} STM topography of \ang{31} twisted $\ch{CrCl3}^{\beta}$ domain. Tunneling parameters: $V=\qty{1.8}{\volt}$, $I=\qty{200}{\pico\ampere}$. \textbf{(b)} Reciprocal lattice construction of \ch{CrCl3} and Au. \textbf{(c)},\textbf{(e)} FT of $\mathrm{d}I/ \mathrm{d}V$ map at \qty{.33}{\electronvolt}, having distinct moiré components (green, red, pink). \textbf{(d)} Energy dependence of FT intensities extracted from the $\mathrm{d}I/ \mathrm{d}V$ maps for different moiré components.}\label{fig4}
\end{figure}

We now show that the observed moiré patterns originate from electronic effects rather than from any modulation of the out-of-plane atomic positions of \ch{CrCl3}. STM topographs mix contributions from electronic states over a broad energy window and may therefore superimpose several moiré modulations of electronic origin. To overcome this limitation, we use tunneling spectroscopy to isolate the local density of states at a well-defined energy, enabling each moiré component to be individually identified in Fourier space.  To this end, we performed scanning tunneling spectroscopy over a $\qty{120}{\nano\meter}\times\qty{120}{\nano\meter}$ region (STM topography in Figure~\ref{fig4}a), where the \ch{CrCl3} lattice is twisted by \ang{31} relative to the <1$\overline{1}$0>$_\text{Au}$ direction. Differential conductance ($\mathrm{d}I/\mathrm{d}V$) maps were acquired across an energy range from \qty{-0.1}{\electronvolt} to \qty{1.8}{\electronvolt}. The FT of the $\mathrm{d}I/\mathrm{d}V$ map at \qty{0.33}{\electronvolt} (Figure~\ref{fig4}c) reveals three sets of moiré spots, highlighted in green, red, and pink in Figure~\ref{fig4}e. These results demonstrate that the moiré patterns in the \ch{CrCl3}$^{\beta}$ islands manifest not only in topographic images but also in the spectroscopic $\mathrm{d}I/\mathrm{d}V$ signal.

To examine how the spectral weight of each moiré component evolves with energy, we quantified the FT intensity at the corresponding moiré spot positions for each bias voltage. The resulting energy dependence is shown in Figure~\ref{fig4}d. The spectral weights of the three moiré patterns vary strongly with energy. At \qty{0.33}{\electronvolt}, the energy at which the FT in Figure~\ref{fig4}e was obtained, the pink moiré, originating from the conjunction of (6,1)${\ch{CrCl3}}$ and (3,2)${\ch{Au}}$, dominates, despite being absent in the topographic channel. Each moiré pattern exhibits distinct resonances at specific energies, indicating a purely electronic origin. A similar conclusion was previously reached for \ch{CrCl3}$^{\alpha}$ islands \cite{gambari_higher_2024} where a single moiré was reported. In contrast, for the twisted \ch{CrCl3}$^{\beta}$ islands, multiple moiré components coexist, each displaying a distinct energy dependence, further reinforcing our interpretation of their purely electronic origin.

As a working hypothesis, we propose that when the wave vectors of the \ch{CrCl3} electronic Bloch states match the moiré periodicity, a nesting-like effect may occur, leading to a spatial modulation of the electronic wave function. Density Functional Theory (DFT) calculations could provide further insight into this intriguing phenomenon. However, performing DFT calculations for this system is beyond the reach of the present study, primarily because the multiple coexisting moiré periods require an extremely large real-space supercell to be treated explicitly. 

\newpage

To summarize, we investigated monolayer \ch{CrCl3} islands deposited on Au(111). The \ch{CrCl3} lattice is rotated by approximately \ang{30} relative to the close-packed directions of Au(111). The incommensurability between \ch{CrCl3} and Au(111) lattices gives rise to a quasiperiodic pattern resulting from the superposition of several moiré orders. We show that multiple moirés can be observed directly for some orientations, and even when one appears dominantly in topography, additional moiré orders emerge in reciprocal space. Importantly, the spectral weight of these distinct moiré components varies considerably with electron energy, demonstrating that the observed quasiperiodic pattern arises from an electronic, rather than structural, modulation.

When simulating electronic or magnetic properties of moiré systems, it is quite generic to assume a single moiré periodicity in the form of a supercell. For instance, this results inthe preditcion of a non-collinear magnetic order with a well-defined period \cite{fumega_moire-driven_2023} that is driven by the considered moiré, considering additionnal moiré would certainly result in a more complex spin texture. Our results suggest that, in practice, multiple magnetic periods are likely to coexist. This would give rise to a quasiperiodic spin texture that could be probed by spin-resolved STM for instance. Similarly, in systems where mechanisms such as chiral topological superconductivity are associated with a specific moiré period \cite{Kezilebieke_moire_2022}, it becomes natural to ask how the electronic properties evolve when additional moiré periodicities are considered. The coexistence of multiple moiré patterns may therefore lead to a richer and more complex phenomenology than previously anticipated.

\section{Experimental}

All experiments were performed under ultra high vacuum (UHV) conditions. The Au(111) single crystal was cleaned by repeated cycles of argon-ion sputtering and annealing at \qty{450}{\degreeCelsius}. Anhydrous \ch{CrCl3} powder was carefully degassed prior to deposition and evaporated from a Knudsen cell onto the clean Au(111) surface at room temperature. To obtain flat \ch{CrCl3} islands, the sample was subsequently annealed at \qty{150}{\degreeCelsius} for a few minutes.

STM topography measurements were carried out at \qty{4.2}{\kelvin} in UHV using electrochemically etched tungsten tips in both a home-built low temperature STM and a commercial Omicron LT-STM. Scanning tunneling spectroscopy (STS) measurements were performed in situ using the home-built setup. The tunneling setpoints were typically \qty{200}{\pico\ampere} and \qty{1.8}{\volt}. 

The spectroscopic grid (Figure~\ref{fig4}) was recorded over an area of $\qty{120}{\nano\meter}\times\qty{120}{\nano\meter}$, with the STM tip positioned every \qty{234}{\pico\meter} to acquire an $I(V)$ spectrum between \qty{-0.1}{\volt} and \qty{1.8}{\volt}. Differential conductance curves ($\mathrm{d}I/\mathrm{d}V$) were obtained by numerical differentiation of the raw $I(V)$ data. Individual $\mathrm{d}I/\mathrm{d}V$ maps were extracted at each bias voltage and Fourier transformed. The spectral weight of each moiré spot was then quantified as a function of the bias voltage.

\bibliography{achemso-demo}

\begin{acknowledgement}

The authors thanks Pascal David for technical support. This work was supported by the French Agence Nationale de la Recherche through the contract ANR GINET2-0 (ANR-20-CE42-0011) and the ANR MASCOTE (ANR-24-CE30-1342). E.G. acknowledged the GDR-NS-CPU for funding it research stay in IPCMS. 

\end{acknowledgement}

\begin{suppinfo}

The following file is available free of charge.
\begin{itemize}
  \item Supporting Information
\end{itemize}

\end{suppinfo}


\end{document}